\documentclass[
    prd,
    noeprint,
    nofootinbib,
    reprint,
    superscriptaddress,
]{revtex4-2}

\usepackage[T1]{fontenc}
\usepackage{amsmath,amssymb}
\usepackage{mathtools}
\usepackage[mathcal]{euscript}
\usepackage{microtype}
\usepackage{bm}
\usepackage{subfigure}

\usepackage{multirow}
\usepackage{graphicx}
\graphicspath{{figures/}}
\usepackage{wrapfig}

\usepackage[svgnames,table]{xcolor}
\usepackage[
    colorlinks=true,
    citecolor=DimGray,
    linkcolor=SlateBlue,
    urlcolor=Navy,
    ]{hyperref}

\usepackage[capitalise]{cleveref}
\usepackage[
    range-phrase=\text{\textendash},
    range-units=single,
]{siunitx}
\DeclareSIUnit{\year}{yr}
\usepackage{dcolumn}

\usepackage{comment}
\usepackage{orcidlink}

\newcommand{\sechead}[1]{\vspace*{10pt}
\noindent\emph{#1 ---}
}

\begin{document}

\title{Low-loss amorphous dielectric metasurface mirror for precision optical cavities}

\author{Swadha Pandey\,\orcidlink{0000-0002-2426-6781}}
\email{swadha@mit.edu}
\author{Peter Fritschel}
\author{Matthew Evans\,\orcidlink{0000-0001-8459-4499}}
\affiliation{LIGO Laboratory, Department of Physics, Massachusetts Institute of Technology, Cambridge, MA 02139, USA}

\date{\today}

\begin{abstract}
We report the design and experimental realization of a low-loss dielectric metasurface mirror operating at \qty{1064}{\nm}, the primary wavelength of interferometric gravitational-wave detectors. 
The mirror consists of a single layer of titanium dioxide nanopillars on a glass substrate, achieving high reflectivity via guided-mode resonance.
The fabricated mirror has a peak reflectance of 99.10\% with a total optical loss of 0.26\%, representing, to our knowledge, the highest reflectance and lowest optical loss experimentally demonstrated in a substrate-supported purely amorphous dielectric metasurface mirror operating in the visible to near-infrared.
This single-layer metasurface is projected to reduce coating Brownian noise by a factor of 3--5 relative to conventional dielectric Bragg mirrors at comparable reflectivity, which would improve the performance of precision optical cavities currently limited by this noise. 
These results indicate the potential of dielectric metasurfaces as a platform for next-generation precision measurement systems.
\end{abstract}

\maketitle

From gravitational-wave detectors to optical atomic clocks, cavity-based precision experiments are increasingly limited by coating Brownian noise in the cavity mirrors~\cite{Levin:1998, Harry:2006}. These mirrors are typically realized as dielectric Bragg stacks composed of tens of alternating high- and low-index quarter-wavelength layers. Mechanical dissipation within the coating materials gives rise to thermally driven fluctuations, and the resulting Brownian noise scales with both the total thickness and mechanical loss of the multilayer stack~\cite{Levin:1998, Harry:2002}. As a result, coating Brownian noise has emerged as a fundamental limitation in state-of-the-art optical cavities~\cite{Matei:2017, Lee:2026}, and extensive efforts have focused on identifying coating materials with intrinsically lower mechanical loss~\cite{Cole:2013, Vajente:2021, McGhee:2023}.
 
Metasurface mirrors achieve high reflectivity using a single layer of sub-wavelength nanostructures via guided-mode resonances engineered using a periodic structure~\cite{Magnusson:1992, Wang:1993}. This enables reflectivities comparable to conventional Bragg stacks, but using an order of magnitude less coating material, providing a pathway towards substantially lower Brownian noise mirrors~\cite{Heinert:2013, Dickmann:2018}. High-reflectivity nanostructure mirrors based on crystalline silicon have demonstrated reflectivities exceeding 99\%~\cite{Bruckner:2010} in the telecommunications band, and have recently been incorporated into high finesse laser cavities~\cite{Dickmann:2023}. However, many precision measurement systems, including gravitational-wave detectors and optical reference cavities, operate in the visible to near-infrared region outside the telecommunications band, particularly near \qty{1064}{nm}~\cite{LIGO:2016}. Resonant mirrors have been used to achieve high reflectivities in this wavelength range using partially~\cite{Bruckner:2008} or purely~\cite{Atikian:2022} crystalline materials. Nevertheless, amorphous dielectric materials offer practical advantages for precision optics, including compatibility with conventional substrates (such as fused silica) and the absence of intrinsic crystalline birefringence that can introduce excess noise~\cite{Yu:2023}. Achieving high-reflectivity, sub-percent loss metasurface mirrors in this wavelength range using purely amorphous materials has remained a significant challenge.

In this paper, we report the design and experimental realization of a sub-percent loss dielectric metasurface mirror operating at \qty{1064}{nm}. The device, which consists of a single layer of titanium dioxide nanostructures on a glass substrate, achieves a peak  reflectance of $99.10\% \pm 0.04\%$ with transmittance $0.64\% \pm 0.04\%$ and inferred optical loss of $0.26\% \pm 0.05\%$ at wavelength \qty{1064.3}{\nm}. Projections of coating Brownian noise indicate a factor of 3--5 improvement relative to Bragg stacks made of amorphous materials with comparable reflectance. To our knowledge, this is the highest reflectance and lowest optical loss experimentally demonstrated in a substrate-supported purely amorphous dielectric metasurface mirror operating in the visible to near-infrared, providing a pathway toward single-layer mirrors with both low optical loss and low coating Brownian noise for next-generation precision measurement systems.

While this work focuses on \qty{1064}{\nm} for ground-based gravitational-wave detectors and ultrastable optical cavities, the design and fabrication techniques are broadly applicable across the visible to near-infrared, where low-noise metasurface mirrors could benefit optical atomic clocks operating at visible wavelengths~\cite{Ludlow:2015} and space-based gravitational-wave detectors~\cite{LISA:2017}.

\begin{figure}[]
    \centering
    \includegraphics[width=0.9\columnwidth]{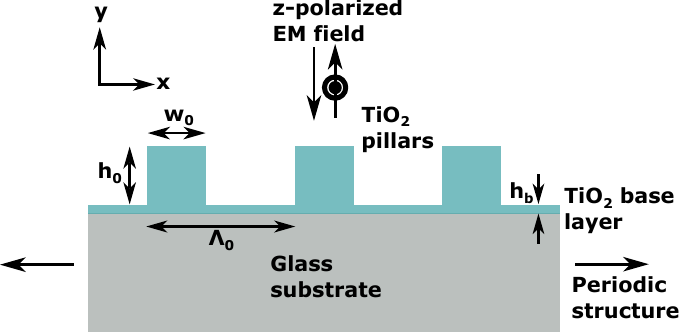}
    \caption{Metasurface mirror consisting of a periodic array of rectangular titanium dioxide (TiO$_2$) nanopillars on a TiO$_2$ base layer of thickness $h_b$, which is on a glass substrate. The nanopillars have width $w_0$, height $h_0$, and periodicity $\Lambda_0$. The structure is periodic in the x-direction and uniform in z-direction, and is optimized to reflect z-polarized light at the target wavelength.}
    \label{fig:concept}
\end{figure}

\sechead{Design}
The metasurface mirror consists of a one-dimensional periodic array of rectangular titanium dioxide (TiO$_2$) nanopillars fabricated on a TiO$_2$ base layer of thickness $h_b$, fabricated on a glass substrate, as shown in \cref{fig:concept}. The nanopillars have width $w_0$, height $h_0$, and period $\Lambda_0$. These geometric parameters are optimized to maximize the reflectivity for normally incident transverse electric (TE) polarized light at the target wavelength $\lambda_0=\qty{1064}{nm}$. The optimization procedure,  implemented using finite-difference time-domain (FDTD) simulations, is described in the Supplementary Information.

\begin{table}
    \centering
    \caption{Parameter values: pillar width ($w_0$), pillar height ($h_0$), nanopillar periodicity ($\Lambda_0$), and base layer height ($h_b$), as shown in \cref{fig:concept}. Values are shown for the optimized and fabricated mirror designs along with loss-free simulated transmittance T$_\mathrm{sim}$ at $\lambda_0=$\qty{1064}{\nm}.}
    \label{tab:params}
    \begin{tabular}{|c || c | c | c |} 
        \hline
        Parameter & Optimized & Fabricated & Unit\\ 
        \hline\hline
        $w_0$ & 218 & 219 & nm\\
        \hline
        $h_0$ & 212 & 213 & nm\\ 
        \hline
        $\Lambda_0$ & 730 & 730 & nm\\
        \hline
        $h_b$ & 15 & 15 & nm\\ 
        \hline
        T$_\mathrm{sim}$ & \qty{0.3}{ppm} & \qty{52}{ppm} & -\\ 
        \hline
    \end{tabular}
\end{table}

\begin{figure}[]
    \centering
    \includegraphics[width=0.9\columnwidth]{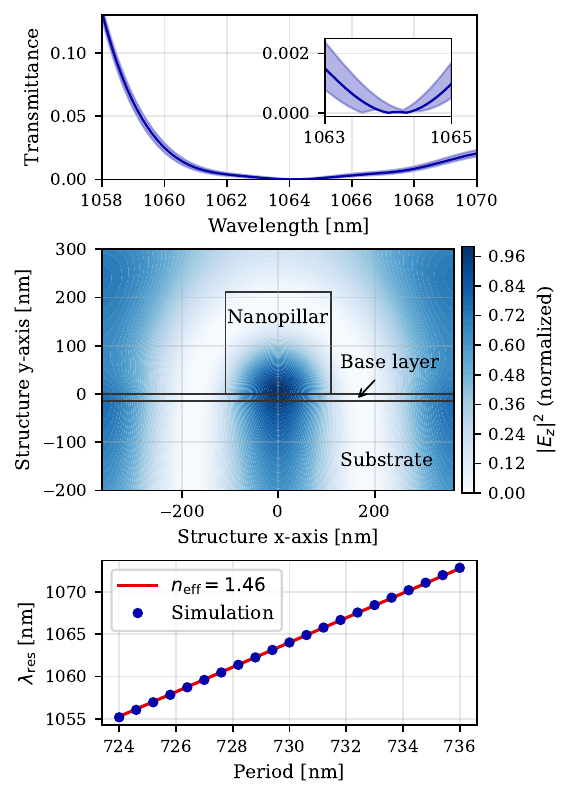}
    \caption{Simulation results for optimized mirror dimensions. \emph{Top:} transmittance spectrum with shaded region showing $\pm$\qty{1}{\nm} variation in $w_0$ and $\pm$\qty{3}{\nm} in $h_0$, independently and in combination, with inset showing the low-transmittance region near resonance at \qty{1064}{\nm}. \emph{Middle:} $\rvert E_z\lvert^2$ field profile at $\lambda_0=$\qty{1064}{nm} on resonance. \emph{Bottom:} resonance wavelength $\lambda_{\mathrm{res}}$ vs. device period $\Lambda_0$, with linear fit yielding $n_{\mathrm{eff}}=1.46$.}
    \label{fig:mode}
\end{figure}

The optimized device parameters are listed in \cref{tab:params}. Numerical simulations predict a transmittance of \qty{0.3}{ppm} at $\lambda_0=\qty{1064}{\nm}$ using a total nanostructured layer thickness of only \qty{227}{\nm}. For comparison, a conventional dielectric Bragg mirror composed of the same materials needs $\sim$\qty{5}{\micro\meter} of coating thickness to achieve the same transmittance. The fabricated device parameters are also shown in \cref{tab:params}, and differ slightly from the nominal design due to fabrication tolerances.

The simulated transmittance spectrum for the optimized device dimensions is shown in \cref{fig:mode}. The shaded region shows the envelope of spectra obtained by varying $w_0$ by $\pm$\qty{1}{\nm} and $h_0$ by $\pm$\qty{3}{\nm}, independently and in combination, demonstrating robustness to fabrication-induced dimensional variations. $\Lambda_0$ and $h_b$ are not varied as they are well-controlled through fabrication. These simulations assume loss-free dielectric materials and do not include absorption or roughness-induced scattering losses.

The corresponding TE field distribution at the resonance wavelength $\lambda_0=\qty{1064}{nm}$ is also shown in \cref{fig:mode}, illustrating the formation of a resonant mode within the nanostructure.

The high reflectivity arises from a guided-mode resonance supported by the periodic nanostructure layer~\cite{Magnusson:1992, Wang:1993}. In this mechanism, incident light couples to a guided mode of the effective waveguide formed by the combined nanostructured TiO$_2$ layer and base layer, resulting in resonant enhancement of reflectivity. To verify this interpretation, we compute the resonance wavelength $\lambda_{\mathrm{res}}$ as a function of the device period $\Lambda_0$. As shown in \cref{fig:mode}, the resonance wavelength varies linearly with period, consistent with the first-order GMR phase-matching condition~\cite{Wang:1993}
\begin{equation}
\lambda_{\mathrm{res}} = n_{\mathrm{eff}} \Lambda_0,
\end{equation}
where $n_{\mathrm{eff}}$ is the effective index of the guided mode. A linear fit yields $n_{\mathrm{eff}}=1.46$. The transmittance spectrum shown in \cref{fig:mode} exhibits near-zero transmittance on resonance ($T_{\mathrm{min}}<10^{-6}$), verified using rigorous coupled-wave analysis (RCWA)~\cite{Moharam:1981}.

The base layer increases the effective refractive index of the waveguide, allowing the resonance condition to be satisfied with shorter and narrower nanopillars than would be required in a pillar-only design, where all mode confinement must be provided by the pillars alone. This reduction in pillar height and width relaxes the demands on etch depth and deposition time respectively, significantly simplifying the fabrication process.

The design and fabrication presented in this work targets a single linear polarization (TE) for a one-dimensional periodic metasurface array, but can be extended to polarization-independent operation using a two-dimensional array.

\begin{figure}[]
    \centering
    \begin{subfigure}
        \centering
        \includegraphics[width=0.8\columnwidth]{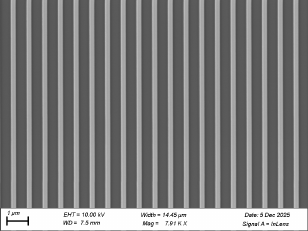}
    \end{subfigure}
    \begin{subfigure}
        \centering
        \includegraphics[width=0.8\columnwidth]{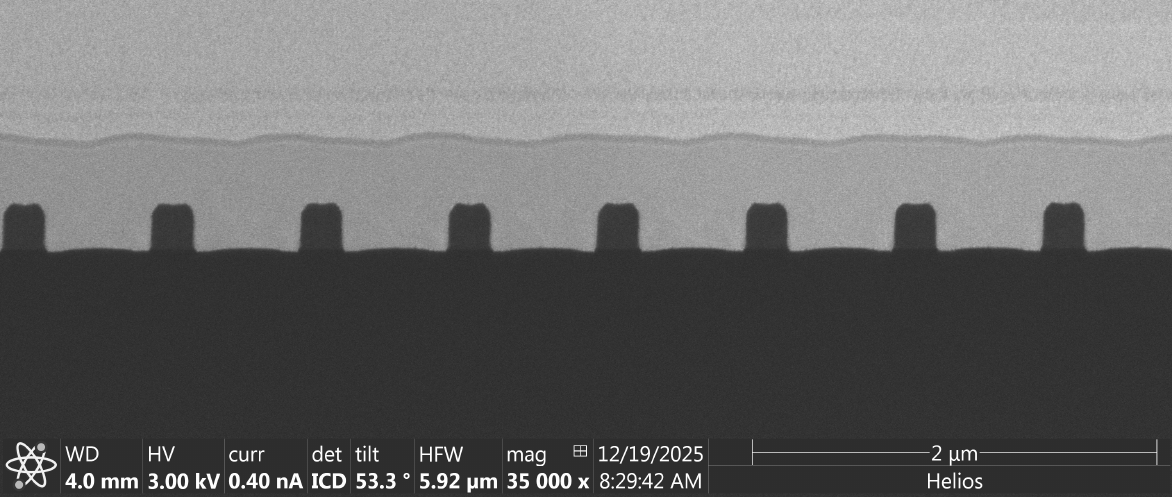}
    \end{subfigure}
    \caption{Scanning electron microscope (SEM) images of the fabricated metasurface mirror. \emph{Top:} top-view image showing periodic TiO$_2$ nanopillar array. \emph{Bottom:} cross-sectional image obtained via focused ion beam (FIB) milling.}
    \label{fig:sem}
\end{figure}

\sechead{Fabrication and measurement}
The device was fabricated on a \qty{1}{inch} diameter fused silica substrate using electron-beam lithography, atomic layer deposition (ALD) of TiO$_2$, and reactive-ion etching, with patterned areas of \qty{500}{\micro\meter} $\times$ \qty{500}{\micro\meter}. The TiO$_2$ ALD was done at \qty{100}{\celsius} to ensure an amorphous film; the complete fabrication flow is described in the Supplementary Information. The fabricated structure was characterized using scanning electron microscopy (SEM), including cross-sectional imaging via focused ion beam (FIB) milling, with representative top and side view images shown in \cref{fig:sem}. The SEM images show uniform periodicity and dimensional control within the imaged regions, and the pillar width ($w_0$) and periodicity ($\Lambda_0$) were measured from the SEM data. The cross-sectional images show smooth, near-vertical sidewalls enabled by the ALD TiO$_2$ process~\cite{Devlin:2016}, which reduces scattering losses from sidewall roughness. The pillar height ($h_0$) and base layer thickness ($h_b$) were measured using atomic force microscopy (AFM) and ellipsometry, respectively. The resulting device dimensions are summarized in \cref{tab:params}.

\begin{figure}[]
    \centering
    \includegraphics[width=0.8\columnwidth]{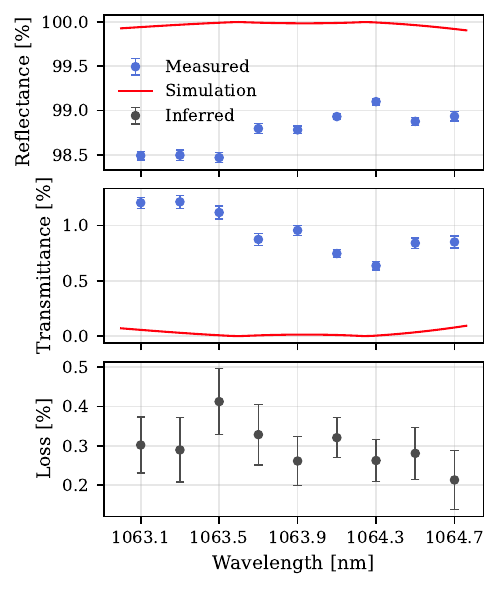}
    \caption{Measured (blue) and inferred (gray) optical performance of the fabricated mirror alongside loss-free numerical simulation (red): reflectance (top); transmittance (middle); inferred optical loss (bottom).}
    \label{fig:meas}
\end{figure}

The optical performance was characterized using a tunable laser source spanning \qtyrange{1063.1}{1064.7}{nm}. Measurements were performed using a Gaussian beam with radius (1/e$^2$ intensity) of \qty{50}{\micro\meter}, ensuring that the beam sampled a uniform region well within the patterned area to avoid edge effects. The reflectance ($R$) and transmittance ($T$) were measured as a function of wavelength, and the total optical loss ($L$, absorption and uncollected scatter) was inferred from energy conservation as $L=1-R-T$. The measured reflectance, transmittance, and inferred loss spectra are shown in \cref{fig:meas}, together with loss-free numerical simulations. The device exhibits a peak reflectance of $(99.10 \pm 0.04)\%$ at the resonance wavelength \qty{1064.3}{nm}, with corresponding transmittance and total optical loss of $(0.64 \pm 0.04)\%$ and $(0.26 \pm 0.05)\%$ respectively. The measured transmittance is much larger than the simulated value for the fabricated dimensions (\qty{52}{ppm}), likely due to the finite angular spread of the \qty{50}{\micro\meter} probe beam as well as shape imperfections and non-uniformity within the illuminated region. These measurements demonstrate high reflectivity with low optical loss in a single-layer amorphous dielectric metasurface mirror, though reaching ppm-level losses requires the improvements discussed below.

\begin{figure}[]
    \centering
    \includegraphics[width=0.9\columnwidth]{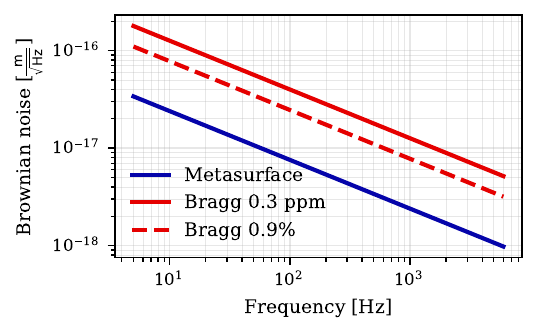}
    \caption{Projected coating Brownian noise for the designed metasurface mirror (blue), compared to that from Bragg mirrors of transmittance \qty{0.3}{ppm} (red, solid) to match the ideal metasurface design, and transmittance 0.9\% (red, dashed) to match the experimentally measured reflectance.}
    \label{fig:noise}
\end{figure}

\sechead{Brownian noise}
We calculate the expected coating Brownian noise reduction from a Bragg stack to a metasurface mirror composed of the same materials and having equivalent reflectance.
We first compute~\cite{Hong:2013,gwinc} the coating Brownian noise of two Bragg mirrors, both composed of alternating quarter-wave layers of TiO$_2$ and SiO$_2$, that have transmittance values of \qty{0.3}{ppm} and 0.9\%, to match the respective design and experimentally measured reflectance values of the metasurface mirror.
Using the direct Levin approach~\cite{Levin:1998} adapted for nano-scale metasurface geometries~\cite{Kroker:2017,Pandey:2026}, we also evaluate the expected coating Brownian noise of a mirror that uses the metasurface design presented in this paper for its reflective surface. 
Following Ref.~\cite{Pandey:2026}, the pressures obtained from the Maxwell stress tensor at each interface are decomposed into a period-averaged part, whose response is computed for a coated half-space~\cite{Hong:2013} with the nanostructure layer homogenized, and a zero-mean periodic part, computed by finite-element analysis of a unit cell.
The two responses are combined coherently, including their cross term.
This captures both the beam-scale and the nanostructure-scale response of the coating.
The displacement noise amplitude spectral densities are computed at 300 K for a Gaussian beam of radius \qty{50}{\micro\meter} at the mirror, corresponding to that of the optical measurement configuration. 
Material parameters corresponding to TiO$_2$~\cite{Flaminio:2010} and SiO$_2$~\cite{gwinc} were used in the calculation, and are displayed in \cref{tab:brownian}. 
The refractive indices of these materials were determined via spectroscopic ellipsometry.
The \qty{0.3}{ppm} and 0.9\% transmittance Bragg mirrors have coatings of thickness \qty{5.3}{\micro\meter} and \qty{2.0}{\micro\meter} respectively, about an order of magnitude more than that of the $\sim$\qty{230}{\nm} metasurface. 
As shown in \cref{fig:noise}, the metasurface mirror reduces the projected coating Brownian noise at \qty{100}{\Hz} by factors of 5.3 and 3.2 relative to the \qty{0.3}{ppm} and 0.9\% transmittance Bragg mirrors respectively. 
Fabricating the metasurface using Ta$_2$O$_5$-TiO$_2$ would further reduce coating Brownian noise, given its order-of-magnitude lower mechanical loss of $\phi=3.9\times10^{-4}$~\cite{Granata:2020}.

\begin{table}
    \centering
    \caption{Values of mechanical loss ($\phi$), Young's modulus (Y), Poisson's ratio ($\nu$), and refractive index at a wavelength of \qty{1064}{\nm} (n), for TiO$_2$ and SiO$_2$.}
    \label{tab:brownian}
    \begin{tabular}{|c || c | c | c |} 
        \hline
        Property & TiO$_2$ & SiO$_2$ & Unit\\ 
        \hline\hline
        $\phi$ & $6.3\times10^{-3}$ & $2.3\times10^{-5}$ & -\\
        \hline
        Y & 290 & 70 & GPa\\ 
        \hline
        $\nu$ & 0.28 & 0.19 & -\\
        \hline
        n & 2.34 & 1.45 & -\\ 
        \hline
    \end{tabular}
\end{table}

\sechead{Future pathway}
We have demonstrated a low-loss metasurface mirror with reflectance exceeding 99\% and sub-percent optical loss at \qty{1064}{nm}, achieved using a single TiO$_2$ nanopillar layer of thickness $\sim$\qty{230}{nm}, an order of magnitude thinner than conventional dielectric Bragg coatings of comparable reflectance. This thinner coating reduces the projected coating Brownian noise by a factor of 3--5, as shown in \cref{fig:noise}. 

Numerical simulations indicate that further optimization of the nanostructure geometry could enable reflectivities exceeding 99.9\% with reduced sensitivity to dimensional variation. For the present geometry, improvements in reflectivity and loss will require further optimization of fabrication processes and materials. Optical loss in the present device is attributed to both scattering from structural imperfections and material absorption of the ALD TiO$_2$. Scattering losses can be reduced through improved etch processes and dimensional control to minimize surface roughness and height variation, as well as improved deposition to reduce imperfections in nanostructure shape. Absorption in TiO$_2$ films deposited by ALD can be reduced through post-deposition treatment and further optimization of deposition parameters. 

In addition, scaling the metasurface fabrication to larger lateral dimensions while maintaining uniformity will be necessary for implementation in large-scale precision optical cavities, especially gravitational-wave detectors. This will require controlling dimensional variations to the nanometer level across apertures of tens of centimeters, which motivates replacing electron-beam lithography with large-area techniques such as photolithography and interference lithography. Hybrid designs combining a metasurface with a reduced Bragg stack have also been implemented as a route to higher finesse optical cavities while preserving the thermal noise advantage~\cite{Dickmann:2023}.

The demonstrated combination of high reflectivity, sub-percent optical loss, and an order-of-magnitude reduction in coating thickness, using conventional substrates and fabrication processes, represents a step toward amorphous metasurface mirrors with reduced coating Brownian noise for precision optical systems such as gravitational-wave detectors and ultrastable optical cavities.

\sechead{Acknowledgments}
SP thanks Mac Hathaway for insightful discussions on the fabrication process, Stephan Kraemer for support with imaging, and Sachin Vaidya for helpful discussions on analysis. This work was partially supported by the LIGO Laboratory, which is funded by the U.S. National Science Foundation award PHY-2309200. This work was carried out in part through the use of MIT.nano's facilities. This work was performed in part at the Harvard University Center for Nanoscale Systems (CNS); a member of the National Nanotechnology Coordinated Infrastructure Network (NNCI), which is supported by the National Science Foundation under NSF award no. ECCS-2025158.

\bibliography{mm}

\end{document}